\documentclass[aps,pra,twocolumn,showpacs,superscriptaddress,groupedaddress]{revtex4}

\usepackage{amsmath, hyperref, color, braket, amssymb, natbib, graphicx, tabularx,float}

\DeclareMathOperator{\tr}{tr}
\newcommand{\refeq}[1]{Eq.~(\ref{#1})}

\begin{document}

\title{Quantum State Reconstruction of an Oscillator Network in an Optomechanical Setting}
\author{Darren W. Moore}
\email{dmoore32@qub.ac.uk}
\affiliation{School of Mathematics and Physics, Queen's University Belfast, BT7 1NN, UK}
\author{Tommaso Tufarelli}
\email{tommaso.tufarelli@nottingham.ac.uk}
\affiliation{School of Mathematical Sciences, University of Nottingham, Nottingham NG7 2RD, United Kingdom}
\author{Mauro Paternostro}
\email{m.paternostro@qub.ac.uk}
\affiliation{School of Mathematics and Physics, Queen's University Belfast, BT7 1NN, UK}
\author{Alessandro Ferraro}
\email{a.ferraro@qub.ac.uk}
\affiliation{School of Mathematics and Physics, Queen's University Belfast, BT7 1NN, UK}

\begin{abstract}
We introduce a scheme to reconstruct an arbitrary quantum state of a mechanical oscillator network. We assume that a single element of the network is coupled to a cavity field via a linearized optomechanical interaction, whose time dependence is controlled by a classical driving field. By designing a suitable interaction profile, we show how the statistics of an arbitrary mechanical quadrature can be encoded in the cavity field, which can then be measured. We discuss the important special case of Gaussian state reconstruction, and study numerically the effectiveness of our scheme for a finite number of measurements. Finally, we speculate on possible routes to extend our ideas to the regime of single-photon optomechanics.

\end{abstract}

\maketitle

\section{Introduction} %%%%%%%%%%INTRODUCTION
Quantum optomechanics exploits radiation pressure to couple photons and mechanical oscillators. The field has progressed significantly in the last decade and is now entering a promising stage where the observation of quantum effects in macroscopic objects appears to be within grasp \cite{meystre2013short, aspelmeyer2014cavity, rogers2014hybrid}. Substantial theoretical and experimental effort has been put into the preparation of mechanical systems, typically consisting of vibrating mirrors or membranes, in interesting non-classical states. In such a context, an important question arises: how can we verify that the mechanical state prepared in an experiment is indeed the desired one? The design of successful strategies to achieve such verifications requires the experimental estimation of the density operator of a mechanical system. However, it is well known that the full information encoded in the density operator cannot be accessed through the measurement of a single observable. %To experimentally reconstruct the density operator 
One must instead collect the measurement statistics of several distinct observables, a task which requires access to many copies of the quantum system of interest. These could be obtained, for instance, by repeating the same experiment with the same initial conditions. By post-processing the outcomes of such measurements, an experimentalist can estimate the density operator via techniques known as \textit{quantum tomography} and \textit{quantum state reconstruction} \cite{ParisRehacek,LvovskyRaymer}. Perhaps the best known example in this context is the reconstruction of the Wigner function of an oscillator (which brings about an amount of information equivalent to that of the density operator) through a Radon transform of the quadrature probability densities \cite{Smithey93}. %(we recall that a quadrature operator is a linear combination of position and momentum operators.).

In an optomechanical setting various approaches to quantum state reconstruction have been explored in the literature \cite{Braginsky:95,aspelmeyer2014cavity,vanner2015towards}, in particular employing weak or quantum non-demolition measurements of mechanical quadratures \cite{caves1981quantum,Braginsky:95,heidmann1996qnd,clerk2008back-action}. Other techniques that have been put forward include the use of short laser pulses to prepare and read out the mechanical state \cite{Braginsky1978, Vanner27092011, NatComm-2295}, the exploitation of a detuned driving field \cite{1367-2630-8-6-107}, and the measurement of the phonon-number operator \cite{Connell:2010}. The possibility of a precise readout has also opened the way towards feedback cooling of a mechanical oscillator \cite{wilson2015measurement}. Very recently, high-efficiency state estimation of a mechanical oscillator through techniques based on Kalman filtering have been implemented experimentally, paving the way to the real-time reconstruction of mechanical state-space configurations, and their quantum-limited control \cite{wieczorek2015optimal}.

\begin{figure}[b!]
\includegraphics[width=0.8\columnwidth]{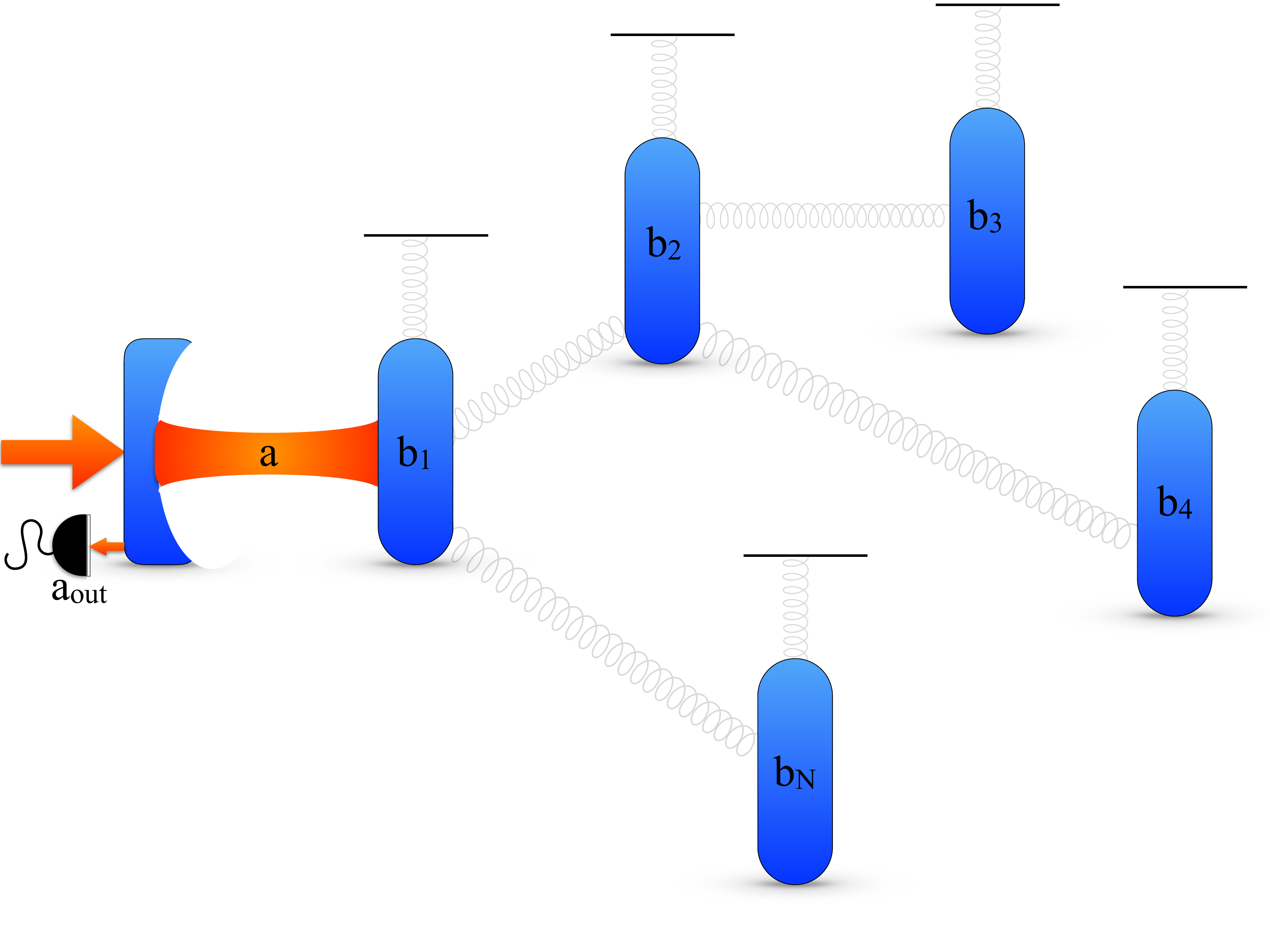}
\caption{Sketch of the system under consideration: $a$ is the annihilation operator of the single intra-cavity mode, whereas $b_1,\dots,b_N$ refer to the mechanical oscillators composing the network. Only the mechanical mode $b_1$ is directly coupled to the radiation field. The driving field (red arrow) can be modulated in order to realize the time-varying linearized radiation-pressure coupling needed to reconstruct a selected quadrature of the oscillator network. Only the field $a_{\rm out}$ leaking from the cavity is eventually measured via homodyne detection. Repeating the protocol for a sufficient set of quadratures, the state of the entire mechanical network can be reconstructed.} \label{f:sketch}
\end{figure}

Besides optomechanical systems comprising a single mechanical oscillator, one can envisage situations in which multiple mechanical oscillators interact in a small network and operate close to the quantum regime \cite{lin2010coherent, massel2012multimode, shkarin2014optically} (Fig.~\ref{f:sketch}). The latter would implement systems of interacting bosons which are of paramount interest in a variety of contexts --- including quantum thermodynamics \cite{gemmer2004quantum} and quantum simulators \cite{johanning2009quantum}, where they represent an ideal playground to test fundamental issues like equilibration \cite{gogolin2016equilibration}, heat transport \cite{michel2005fourier, asadian2013heat, nicacio2015thermal, xuereb2015transport, pigeon2016HarmonicTrajectories}, the definition of temperature \cite{hartmann2004existence, ferraro2012intensive, kliesch2014locality}, and the universal scaling of ground-state entanglement \cite{plenio2005entropy, eisert2010colloquium}. In addition, interacting quantum oscillators have been proposed as a valid route towards quantum computation \cite{aolita2011gapped, menicucci2011graphical, schmidt2012optomechanical, houhou2015generation}. Quantum state reconstruction is instrumental in all these settings. Outside the optomechanical domain, the reconstruction of an oscillator network can be accomplished using a two-level system interacting with one node of the network --- provided the coupling constant is time-dependent and can be controlled by the experimenter \cite{PhysRevA.83.062120,PhysRevA.85.032334}. However, apart from measuring each oscillator individually, to the best of our knowledge no method has been proposed for the efficient readout of the quantum state of an oscillator network in an optomechanical setting.

In this paper we propose a protocol of quantum state reconstruction for the mechanical portion of a generalized optomechanical system, featuring a single high-quality cavity mode coupled to a network of mechanical oscillators (see Fig.~\ref{f:sketch}). Our protocol relies on the so-called linearized radiation pressure interaction, and exploits measurements on the accessible output modes of the optical cavity (rather than the mechanical modes of the oscillator network which are typically challenging to measure directly). By controlling in time the interaction strength, we show that it is possible to encode information about any mechanical quadrature in the cavity light, which can then be measured through the output fields leaking out of the system. Specifically, we discuss how an arbitrary moment of the selected quadrature can be estimated via appropriate light-quadrature measurements, followed by the inversion of a linear system of equations. Our scheme shares an important advantage with Ref.~\cite{PhysRevA.85.032334}: it requires minimal access to the oscillator network, in that it can be probed through interaction with just one of its elements.

The paper is organised as follows. In Section~\ref{sec:oneoscillator} we present our state reconstruction scheme applied to a single mechanical oscillator, a simple example which provides a gentle introduction to the more technical case of a network. We start by showing how the dynamics of interest can be solved analytically, then discuss how the interaction profile can be designed to encode a chosen mechanical quadrature in the cavity light mode. Finally, we conclude the section by explaining how to measure indirectly the light mode through collection of the cavity output field. Section~\ref{sec:network} presents the main results of this paper, and generalizes our reconstruction procedure to a network of mechanical oscillators. We provide sufficient conditions under which the quantum state of the entire network can be reconstructed, as well as an explicit analytical procedure to design the interaction profile. Section~\ref{sec:Gauss} deals with the important special case of Gaussian states, and provides a numerical simulation of our scheme for the realistic case of a finite number of measurements. In Section~\ref{sec:fottosingle} we present some ideas about extending our state reconstruction scheme to the single-photon regime of optomechanics. Finally, we draw our conclusions in Section~\ref{concussioni}.

\section{A single oscillator}\label{sec:oneoscillator} %%%%%%%%%%OPTOMECHANICS SETTING
\subsection{Optomechanical model}
Our starting point is the so-called linearized optomechanical interaction, which involves a cavity mode with annihilation operator $a$ and an oscillating mirror whose excitations are described via a second annihilation operator $b$. The cavity is pumped by a resonant classical field, and in a frame rotating with the cavity frequency the Hamiltonian reads
\begin{equation}
H=\omega_mb^\dagger b+g(t)X(b+b^\dagger)\;, \label{linear_1}
\end{equation}
where $\omega_m$ is the mechanical oscillator frequency, $g(t)$ is a time-dependent coupling constant controlled via the amplitude of the classical driving field, and $X=(a+a^\dagger)/\sqrt{2}$ is a cavity quadrature operator. Notice that an alternative scheme to generate Hamiltonian (\ref{linear_1}) with tunable coupling has been recently reported in \cite{felicetti2016quantum}. Moving to an interaction picture defined by $H_0=\omega_mb^\dagger b$, the Hamiltonian becomes
%%%%%%%%%%%%
\begin{equation}
H_I=g(t)X(be^{-i\omega_mt}+b^\dagger e^{i\omega_mt})\;.
\end{equation}
%%%%%%%%%%%%
The latter satisfies the Schr\"{o}dinger equation $\dot{U}=-iH_IU$, which can be solved via the ansatz $U=e^{i\phi}D(X\beta)$ [with $D(\alpha)\equiv e^{\alpha b^\dagger-\alpha^*b}$ the displacement operator of the mechanical mode], where $\phi$ is assumed to be a time-dependent operator that commutes with the mechanical degree of freedom (namely, $[\phi,b]=[\phi,b^\dagger]=0$), whereas $\beta$ is a time-dependent complex number. Then
\begin{equation}
\dot{U}=i\dot{\phi}U-X(\dot{\beta}\beta^*+\beta\dot{\beta}^*)U-X\dot{\beta}^*Ub\;.
\end{equation}
\noindent Using the unitarity of $U$ and the Baker-Campbell-Hausdorff formula, note that
\begin{align}
\dot{U}U^\dagger &=i\dot{\phi}-X^2(\dot{\beta}\beta^*+\beta\dot{\beta}^*)-X\dot{\beta^*}UbU^\dagger
\;, \\
UbU^\dagger &=b-X\beta^*\;.
\end{align}
\noindent Then the Schr\"{o}dinger equation implies
\begin{align}
i\dot{\phi}+iX^2\text{Im}(\beta\dot{\beta}^*)&+Xb^\dagger\dot{\beta}-b\dot{\beta}^*\\&=-ig(t)Xbe^{-i\omega_mt}+b^\dagger e^{i\omega_mt}\;.
\end{align}
\noindent Matching coefficients produces a set of simultaneous equations:
\begin{equation}
\dot{\beta}=-ig(t)e^{i\omega_mt}\;,~~~~~~~~\dot{\phi}=-X^2\text{Im}(\beta\dot{\beta}^*)\;.
\end{equation}
\noindent For an interaction time $\tau$, one has the solutions
 \begin{align}
\beta&=-i\int_0^\tau g(s)e^{i\omega_ms}ds \label{beta}\;,\\ 
\phi&=-X^2\int_0^\tau\text{Im}(\beta\dot{\beta}^*)ds \label{phi}\;.
\end{align}
%\begin{tabular*}{c c}
%	{\begin{equation}}
%	{\beta=-i\int_0^\tau g(s)e^{i\omega_ms}ds \label{beta}\;,}
%	{\end{equation}} &
%	{\begin{equation}}
%	{\phi=-X^2\int_0^\tau\text{Im}(\beta\dot{\beta}^*)ds \label{phi}\;.}
%	{\end{equation}}
%\end{tabular*}

%\begin{tabular}{l l}
%\begin{equation}\beta=-i\int_0^\tau g(s)e^{i\omega_ms}ds \label{beta}\;,\end{equation} & \begin{equation}\phi=-X^2\int_0^\tau\text{Im}(\beta\dot{\beta}^*)ds \label{phi}\;.\end{equation}
%\end{tabular}

%\begin{minipage}{0.55\linewidth}
%\begin{equation}
%\beta=-i\int_0^\tau g(s)e^{i\omega_ms}ds \label{beta}\;,
%\end{equation}
%\end{minipage}
%\begin{minipage}{0.8\linewidth}
%\begin{equation}
%\phi=-X^2\int_0^\tau\text{Im}(\beta\dot{\beta}^*)ds \label{phi}\;.
%\end{equation}
%\end{minipage}

Finally one can rewrite $U=e^{i\psi X^2}D(X\beta)$ with 
\begin{equation}
\psi=-\int_0^\tau\text{Im}(\beta\dot{\beta}^*)ds\;.\label{psi_1}
\end{equation}
Therefore, the dynamics is described by a mechanical displacement operator whose amplitude depends on the in-phase quadrature operator of the optical mode modified by a quadratic term on the optical mode. 

For the purposes of this reconstruction strategy, the dynamics can be greatly simplified by constructing the interaction profile in such a way as to eliminate the quadratic term (\textit{i.e.}, setting $\psi=0$)\footnote{The measurement of the cavity field in this case is the momentum operator $P$. If $\psi\neq0$ the observable depends on this parameter and is modified to $L=-2\psi X+P$\label{footnote}}.

\subsection{State reconstruction of a single oscillator} \label{sec:1osc}%%%%RECONSTRUCTION OF A SINGLE OSCILLATOR
%To reconstruct the quantum state of the oscillator let us first observe that having the entire sequence of statistical moments of a quadrature suffices to uniquely determine the probability distribution associated with it \comm{this is not true in general, change the sentence}. In the case where only finitely many terms in the sequence are known there exist distribution estimation methods.
{\noindent}We will now show that a link can be established between the cavity quadrature operator $P={i(a^\dagger-a)}/{\sqrt{2}}$ (namely, the canonical momentum operator conjugated to $X$ --- momentum for brevity) and an arbitrary mechanical quadrature. Then, we will use these results to illustrate how the full reconstruction of the oscillator state can be carried out. Let the initial state of light and mechanics be $\rho=\ket{0}\bra{0}\otimes\rho_0$ with $\ket{0}$ the vacuum state of the optical mode (in the displaced frame of reference) and $\rho_0$ the mechanical state to be reconstructed. For the scope of the present discussion it is convenient to switch to the Heisenberg-picture. After an interaction time $\tau$, the cavity field's momentum evolves into
\begin{equation}\label{UPU}
P(\tau)\equiv U^\dagger PU=P-\sqrt{2}|\beta|Q_\theta\;,
\end{equation}
where $Q_\theta\equiv(be^{-i\theta}+b^\dagger e^{i\theta})/\sqrt{2}$ is the arbitrarily chosen mechanical quadrature. The phase of $Q_\theta$ is controlled by the parameter $\beta$, as per $\theta=\arg(\beta)+\frac \pi2$. We rescale the observable as
\begin{equation}\label{probsum}
Q_\text{r}\equiv -\frac{P(\tau)}{\sqrt2|\beta|}=Q_\theta-\frac{P}{\sqrt2 |\beta|},
\end{equation}
from which we can easily deduce a relationship between ${\mathbb P}_\text{r}(q)$, the (measurable) probability distribution of $Q_\text{r}$, and $\mathbb P_\theta(q)$, the probability distribution of $Q_\theta$ in the state $\rho_0$. Such distributions are related by the convolution integral
\begin{equation}\label{convolution}
	{\mathbb P}_\text{r}(q)=\int_{-\infty}^\infty{\rm d}q'\, {\mathbb P}_\theta(q') \, \sqrt{\frac{2}{\pi}}|\beta|e^{-2|\beta|^2(q-q')^2},
\end{equation}
where the Gaussian factor is due to the second term in Eq.~\eqref{probsum} (recall that $P$ exhibits vacuum statistics on our initial state). As the displacement paramenter $|\beta|$ increases, the measured distribution ${\mathbb P}_{\text{r}}(q)$ provides a better approximation to ${\mathbb P}_{\theta}(q)$. In the limit $|\beta|\to \infty$ one would have ${\mathbb P}_\text{r}(q)\to\mathbb P_\theta(q)$. Note however that the magnitude of $|\beta|$ will be limited by physical constraints such as the maximum achievable coupling strength ${\max}_t|g(t)|$ and the requirement to keep interaction times short enough to avoid decoherence. For finite $|\beta|$ it is in principle possible to recover $\mathbb P_\theta(q)$ from ${\mathbb P}_\text{r}(q)$ via standard deconvolution techniques, due to the fact that the Gaussian distribution in Eq.~\eqref{convolution} is fully known. 

{\noindent}As an alternative to the deconvolution approach, one may also exploit Eq.~\eqref{probsum} to establish a relation between the statistical moments of the measured operator $Q_\text{r}$ and the mechanical moments $\braket{Q_\theta^k}$
\begin{align}
\langle Q_\text{r}^n\rangle=%\tr\{(P-\sqrt{2}|\beta|Q_\theta)^n\rho\} \nonumber\\
%&=\tr\Big\{\sum_{k=0}^n\binom{n}{k}P^{n-k}(-\sqrt{2}|\beta|Q_\theta)^k\rho\Big\} \nonumber\\
\sum_{k=0}^n\binom{n}{k}\braket{Q_\theta^{n-k}}\left(\frac{-1}{\sqrt{2}|\beta|}\right)^{k}{\cal V}_k\;,
\label{system1}
\end{align}
where ${\cal V}_k$ indicates the statistical moments of $P$ in the vacuum state, ${\cal V}_k\equiv\bra0P^k\ket0$
\begin{align}\label{vaculo}
{\cal V}_k&=
\left\lbrace\begin{array}{lr}
0 & k \text{ odd}\\
\frac 1{\sqrt\pi}\Gamma\left(\tfrac{k+1}{2}\right)& k \text{ even}
\end{array}\right.,
\end{align}
where $\Gamma$ is the Euler-Gamma function. Eq.~(\ref{system1}) tells us that, having experimentally determined $\braket{Q^j_{\text{r}}}$ ($j=1,\dots,n$), we may calculate $\braket{Q^j_{\theta}}$ ($j=1,\dots,n$) from the data by inverting an $n\times n$ linear system of equations. As discussed in section~\ref{sec:Gauss}, this second approach is particularly convenient when $\rho_0$ is a Gaussian state, in which case the knowledge of first and second moments of $Q_\theta$ (for several values of $\theta$) is sufficient for full state reconstruction.\\

{\noindent}Let us now illustrate the quantum state reconstruction protocol. First, the user selects a quadrature $Q_\theta$ to reconstruct. This choice determines the value of $\arg(\beta)$ as shown above. The modulus of $\beta$ along with the coupling $g(t)$ and interaction time $\tau$ are chosen such that $\psi=0$. This can be accomplished by setting
%\footnote{If $\psi\ne0$, the measurement operator depends on the particular evolution, in the form $L=-2\psi X+P$. Note further that written like this the measurement is not a quadrature operator and must be suitably rescaled if homodyne measurements are to be used.}.  
\begin{equation}\label{g_1mode}
g(t)=\frac{\omega_m}{2\pi}\left(Ae^{-i\omega_mt}+A^*e^{i\omega_mt}+Be^{-2i\omega_mt}+B^*e^{2i\omega_mt}\right)
\end{equation}
\noindent and using Eqs.~(\ref{beta}, \ref{phi}) along with the choice of $\arg(\beta)$ and the condition $\psi=0$ to solve for the coefficients $A$ and $B$. %Notice that imposing $\psi=0$ reduces the unitary evolution to a displacement operation.
After an interaction time $\tau$, the cavity field has assimilated the information from the mechanical mode and is ready to be measured. At this point the coupling is switched off, the measurement is performed, and the result recorded. The system must be reset and the procedure repeated sufficiently many times such that the sampling of the measurement results is reliable. The probability distribution $\mathbb P_\text{r}(q)$ and/or the associated moments may then be estimated from the collected data. Subsequently one may proceed to the deconvolution of $\mathbb P_\text{r}(q)$, [or the inversion of Eqs.~(\ref{system1})] in order to estimate the distribution $\mathbb P_\theta(q)$ (or a finite number of its moments). The procedure must then be repeated for a sufficient number of different quadratures (\textit{i.e.}, different values of $\theta$), such that the state $\rho_0$ may be recovered via the standard inversion techniques of quantum tomography \cite{ParisRehacek,LvovskyRaymer,Smithey93,Vogel}.

\subsection{Measuring the cavity through its output field}
{\noindent}Note that the scheme presented so far relies on the measurement of the intra-cavity field, which is typically not directly accessible. However, by making use of the small but inevitable transmittance of the cavity mirrors, one may measure the associated output fields and infer the intra-cavity field properties via input-output theory \cite{PhysRevA.31.3761}. %Input-output theory relates the input field mode, the output field mode, and the intra-cavity field mode. 
For the case under consideration, assume that the emission rate of the cavity $\kappa$ is small enough as to be negligible during the reconstruction protocol described so far (\emph{i.e.}, $\kappa\tau\ll1$). It is also convenient to assume that such emission occurs through only one of the two cavity mirrors, so that it may be more easily collected. At time $\tau$, we assume that the optomechanical coupling has been switched off, and that the cavity obeys the standard quantum Langevin equation \cite{PhysRevA.31.3761}
\begin{equation}
\dot{a}-\frac \kappa2 a=-\sqrt{\kappa}a_{\text{out}},
\end{equation}
\noindent whose formal solution can be arranged as
\begin{equation}
a(\tau)=e^{-\frac \kappa2 (t_f-\tau)}a(t_f)+\sqrt{\kappa}\int_{\tau}^{t_f}e^{-\frac{\kappa}{2}(t'-\tau)}a_{\text{out}}(t')dt'\;.
\end{equation}
This is a Heisenberg-picture relation indicating that the full information about the cavity field (at the time $\tau$ of interest) is shared between the output field modes and the cavity field at the later time $t_f$. Observe that for $t_f\gg\kappa^{-1}$ the desired information is fully encoded in the output field. Formally, this amounts to the expression
\begin{equation}\label{ideal-readout}
a(\tau)=\sqrt{\kappa}\int_\tau^{\infty}e^{-\frac{\kappa}{2}(t'-\tau)}a_{\text{out}}(t')dt'\equiv f_{\text{out}}
\;,
\end{equation}
where the bosonic operator $f_{\text{out}}$ represents an appropriate combination of output field modes that can be measured directly. Note that, so far, we have considered an idealized cavity in which all internal losses are associated with emission into detectable modes. In the presence of genuine optical losses, which irreversibly deteriorate the amount of accessible information, Eq.~\eqref{ideal-readout} must be modified as follows \cite{1367-2630-14-9-093046, PhysRevLett.112.133605}.
\begin{equation}\label{realistic-readout}
	f_{\text{out}}=\sqrt{1-\epsilon}\, a(\tau)+\sqrt{\epsilon}\,a_{\text{vac}},
\end{equation}
\noindent where $0<\epsilon<1$ is the probability of single-photon loss, while $a_{\text{vac}}$ is a bosonic mode accounting for the associated added noise: it commutes with $f_\text{out}$ and is assumed to display vacuum statistics. Constructing the usual quadrature operators we obtain the relation $P_{\text{out}}=\sqrt{1-\epsilon}\, P(\tau)+\sqrt{\epsilon}\,P_{\text{vac}}$, which can be exploited to obtain relationships analogous to Eqs.~\eqref{convolution} and \eqref{system1}, linking the measurement statistics of $P_{\text{out}}$ to those of $P(\tau)$. In particular, the measured moments read
\begin{align}
\braket{P_\text{out}^n}=&\sum_{k=0}^n\binom{n}{k}(1-\epsilon)^{\frac {n-k}2}\epsilon^{\frac{k}{2}}{\cal V}_k\braket{P^{n-k}(\tau)}\;. \label{realistic-readout-moments}
\end{align}
 The moments $\braket{P_\text{out}^n}$ can be estimated via homodyne detection, so that an inversion of the equations above allows one to retrieve the moments  $\braket{P^{n}(\tau)}$. As outlined in Section \ref{sec:1osc}, the latter can then be rescaled to obtain the moments of $Q_\text{r}$, which in turn allows to retrieve the desired mechanical quadratures. Alternatively, the measured probability distribution of $P_{\text{out}}$ may be deconvoluted to obtain that of $P(\tau)$, if the noise parameter $\epsilon$ is known. Appendix \ref{LossesAppendix} details examples of the reconstruction when the noise parameter is significant. The primary effect is to increase the number of measurements required for a good representation of $\braket{P_{\text{out}}^n}$.

The next significant noise factor is that of damping of the mechanical oscillator at rate $\Gamma$. In the regime in which the reconstruction protocol takes place the mechanical damping rate is very small compared to the cavity decay rate. To clarify, we operate in the resolved sideband regime which requires that $\Gamma\ll\kappa\ll\omega_m$. We thus conclude that the effect of the mechanical damping over the timescale of the interaction with the cavity field is negligible. Furthermore, the explored examples of $g(s)$ (Figs.~\ref{IntProf} and \ref{IntProf2}) required for reconstruction show that the reconstruction is effective over the course of a few mechanical periods.

\section{A network of oscillators}\label{sec:network}
%RECONSTRUCTION OF AN OSCILLATOR NETWORK
\subsection{Hamiltonian and time evolution}
The protocol described above can be generalized to the case of a network of $N$ harmonically coupled oscillators, whose mechanical excitations are described by a set of bosonic operators $b_1,...,b_N$, with $[b_i,b^\dagger_j]=\delta_{ij}$. Only one mechanical oscillator, say $b_1$, is coupled to the optical mode $a$ (see Fig.\ref{f:sketch}). The Hamiltonian for such a system reads $H=H_0+H_{int}\;,$
\begin{align}
H_0&=\sum_n\omega_nb_n^\dagger b_n+\sum_{n<m}J_{nm}(b_nb_m^\dagger +b_n^\dagger b_m)\\&~~~~~~~~~~~~+\sum_{n<m}K_{nm}(b_nb_m+b_n^\dagger b_m^\dagger)\nonumber\;,\\
H_{int}&=g(t)X(b_1+b_1^\dagger)\;, \label{linear_net}
\end{align}
\noindent where $\omega_n$, $J_{nm}$ and $K_{nm}$ are the bare frequencies and coupling constants characterizing the network --- which are assumed to be known in advance \cite{burgarth2011HamiltonianTomography,madalin2016system}. By Williamson's theorem, the mechanical portion of our system can be brought into diagonal form by a symplectic transformation $S$, which has the general structure
\begin{equation}
S=\begin{pmatrix} 
	S_1 & S_2 \\
	S^*_2 & S^*_1
	\end{pmatrix}\;.
\end{equation}
{\noindent}In terms of the mechanical normal modes defined by $S$, the Hamiltonian reads
\begin{align}
&H_0=\sum_n\nu_nd_n^\dagger d_n\;,\\
&H_{int}=g(t)X\sum_n\left(G_nd_n+G^*_nd_n^\dagger\right)\;, \label{Hint}
\end{align}
\noindent where $G_n=(S_1-S_2)^*_{n1}$, $d_n$ are the annihilation operators for the normal modes ($[d_n,d_m^\dagger]=\delta_{nm}$), and $\nu_n$ the associated eigenfrequencies. Moving to the interaction picture defined by $H_0$, one has
\begin{equation}
H_I=g(t)X\sum_jh_j(t)\;,
\end{equation}
\noindent where $h_j=G_jd_je^{-i\nu_jt}+G_j^*d_j^\dagger e^{i\nu_jt}$. These operators have the property that $[h_j(t),h_{j'}(t')]=0$ $\forall j\ne j'$. This allows the unitary for the system to be written as $U=\otimes_ju_j(t)\;,$
\noindent with each $u_j$ satisfying the equation
\begin{equation}
\dot{u}_j=-ih_j(t)u_j(t)\;,
\end{equation}
\noindent with the initial condition $u_j(0)=\mathbb{I}\;.$
\noindent Following Section \ref{sec:oneoscillator}, these equations can be solved by the ansatz
\begin{equation}
u_j=e^{i\phi_j}D(X\beta_j)\;,
\end{equation}
\noindent with $\phi_j$ commuting with all involved mechanical modes. There are two coupled equations associated with each $j$
\begin{align}
\dot{\phi}_j=-iX^2 \text{Im}(\dot{\beta}_j\beta_j^*)\;,~~~~~~~~\dot{\beta}_j=-ig(t)G_j^*e^{i\nu_jt}\;,
\end{align}
\noindent which have solutions
\begin{align}
&\beta_j=-iG_j^*\int_0^tg(s)e^{i\nu_js}ds \label{betas}\;,\\
&\phi_j=-iX^2\int_0^t\text{Im}(\beta_j\dot{\beta}_j^*)ds\;.
\end{align}
\noindent Then 
\begin{equation}
U=e^{i\Psi X^2}D(X\mathbf{\beta})\;,
\end{equation}
\noindent where $\Psi=\sum_j\psi_j$, $\mathbf{\beta}=\begin{pmatrix}
			\beta_1 &
			\beta_2 &
			\dots &
			\beta_N
		\end{pmatrix}^\top$ and

\begin{equation}
\psi_j=-\int_0^t\text{Im}(\beta_j\dot{\beta}_j^*)ds\;.
\end{equation}
\noindent The evolution of this more general system bears a clear resemblance to the single oscillator case. First, it comprises a quadratic term on the optical mode that depends on the global phase $\Psi$. We will set this to zero, extending the method used for the reconstruction of one oscillator (see below). Second, we can recognise an $X$-conditioned multimode displacement on the mechanical modes. In order to proceed with the state reconstruction, it is necessary to introduce two assumptions on the properties of the network. These are 
\begin{itemize}
\item (A1) $G_n\neq0 ~~\forall ~n$.
\item (A2) The spectrum of normal modes $\{\nu_j\}$ is non-degenerate.
\end{itemize}
These assumptions embody the ability of the cavity field to interact with, and distinguish, all the normal modes of the network [see Eq.~(\ref{Hint})]. A further practical requirement is that the interaction time should be sufficiently long to allow the resolution of modes that vibrate with similar frequencies. The dynamics described here is reminiscent of the one derived in Ref.~\cite{PhysRevA.85.032334}, where an oscillator network is probed with an auxiliary two-level system rather than with a cavity field.

\subsection{Quantum state reconstruction}\label{QSRmulti}

Similarly to the single oscillator case, let the initial state of the system be the factorised state $\rho=\ket{0}\bra{0}\otimes\rho_0$ where $\rho_0$ now indicates an arbitrary state of the oscillator network. Reconstruction proceeds as before, with the following modification: there are now multiple mechanical quadratures, defined by $Q_{\theta_j}=(d_je^{-i\theta_j}+d_j^\dagger e^{i\theta_j})/\sqrt2$. Note that these quadratures are defined in terms of the normal mode operators, so that the state is reconstructed in the normal-mode basis. However, a representation in terms of the original modes $b_1,...,b_N$ can be obtained through the inverse symplectic transformation $S^{-1}$, corresponding to a reshaping of the reconstructed Wigner function. As before, let us work in the Heisenberg picture, and let us indicate by $\tau$ the interaction time in which the controlled displacement is implemented. After the interaction, the cavity quadrature $P$ evolves into
\begin{align}\label{Pmulti}
P(\tau)=P-\sqrt{2}\sum_j|\beta_j|Q_{\theta_j},
\end{align}
where $\theta_j=\arg(\beta_j)+\frac \pi2$. By proceeding as for the case of a single mechanical oscillator, one may write a convolution integral connecting the probability distribution of $P(\tau)$ to that of the mechanical quadratures. Since the choice of each quadrature $Q_{\theta_j}$ is determined only by the phase of $\beta_j$, by varying $|\beta_j|$ in Eq.~\eqref{Pmulti} it is possible to measure a sufficient number of linearly independent observables to enable a deconvolution, and hence estimate the multivariate probability distribution of $(Q_{\theta_1},...,Q_{\theta_N})$. As we will be primarily interested in Gaussian states, however, here we focus on the reconstruction of arbitrary moments of the mechanical quadratures, rather than their probability distribution. From Eq.~\eqref{Pmulti}, it follows that the moments of $P$ are linked to the mechanical quadrature moments by
\begin{equation}
\braket{P^n}=\sum\limits_{\{k_j\}}\binom{n}{k_0,k_1,...k_N}{\cal V}_{k_0}\braket{\prod\limits_{1\le j\le N}(-\sqrt{2}|\beta_j|Q_{\theta_j})^{k_j}}\; \label{systemN}
\end{equation}
{\noindent}with the sum over all permutations of integers $(k_1,...,k_N)$ such that $k_1+k_2+...+k_N=n$, and we recall that ${\cal V}_{k_0}\equiv\bra0P^{k_0}\ket0$ is given in Eq.~\eqref{vaculo}. This system of simultaneous equations is under-determined, however we can exploit the freedom in $|\beta_j|$ to generate as many independent extra equations as necessary, involving the same variables but different coefficients. This results in a linear system of equations that, once inverted, provides an arbitrary moment of the mechanical network quadratures. For Gaussian states, we recall that first and second moments will suffice to fully characterize the mechanical quantum state.

As in Section~\ref{sec:1osc}, in order to \textit{(i)} reconstruct an arbitrary quadrature $\braket{Q_{\theta_j}}$ and \textit{(ii)} reduce the dynamics to a multimode displacement (\textit{i.e.} set $\Psi=0$)\footnote{If $\Psi\neq0$ the observable is modified to $L=-2\Psi X+P$\label{footnote}}, it is crucial to properly select the interaction profile $g(s)$. Let us define the latter as
\begin{equation}
g(s)=\frac it\left[\sum_k\left(\frac{g_k}{G_k^*}e^{-i\nu_ks}-\frac{g_k^*}{G_k}e^{i\nu_ks}\right)+he^{-i\omega s}-h^*e^{i\omega s}\right] \label{interN}
\end{equation}
\noindent where the additional term outside the sum does not depend on any frequency of the system (\textit{i.e.} $\omega$ is arbitrary).
\noindent From Eq~(\ref{betas}), it follows that 
\begin{equation}
	\begin{pmatrix}
	\mathbf{\beta}\\
	\mathbf{\beta}^*
	\end{pmatrix}
	=
	{\bf{R}}
	\begin{pmatrix}
	\mathbf{\mathcal{G}}\\
	\mathbf{\mathcal{G}}^*
	\end{pmatrix}
	+\bf{S}\begin{pmatrix}
	h\\
	h^*
	\end{pmatrix}\;,
\end{equation}
\noindent where 
\begin{equation}
\bf{R}=\begin{pmatrix}
	N & M\\
	M^* & N^*
	\end{pmatrix}\;,
\end{equation}
\noindent with matrix elements
\begin{align}
N_{nm}=\frac{G_n^*}{tG_m^*}\int_0^te^{i(\nu_n-\nu_m)s}ds\;,\\
M_{nm}=\frac{G_n^*}{tG_m}\int_0^te^{i(\nu_n+\nu_m)s}ds\;,
\end{align}
\noindent$\mathcal{G}$ is the vector of interaction profile coefficients,
\begin{equation}
\mathbf{\mathcal{G}}=\begin{pmatrix}
					g_1, &
					g_2 ,&
					\dots ,& 
					g_N
					\end{pmatrix}^\top\;,
\end{equation}
\noindent and 
\begin{equation}
\bf{S}=\begin{pmatrix}
		P & Q\\
		P^* & Q^*
		\end{pmatrix}\;,
\end{equation}
where
\begin{align}
&P_n=\frac{G_n^*}{t}\int_0^te^{i(\nu_n-\omega)s}ds\;,\\
&Q_n=\frac{G_n^*}{t}\int_0^te^{i(\nu_n+\omega)s}ds\;,
\end{align}
\noindent with $\bf{S}$ being a rectangular matrix. For long interaction times the matrix $\bf{R}$ is invertible. It can be shown that 
\begin{equation}
\lim_{t\rightarrow\infty} {\bf{R}}={\bf{I}}\Rightarrow\lim_{t\rightarrow\infty}\det({\bf{R}})=1\;.
\end{equation}
This implies that there exists an interaction time such that $\det({\bf{R}})>0$ and hence $\bf{R}$ is invertible. Then, as $\beta$ is chosen, the coefficients $\mathcal{G}$ are determined via
\begin{equation}\label{Glinear}
	\begin{pmatrix}
	\mathcal{G}\\
	\mathcal{G}^*
	\end{pmatrix}
	=
	\bf{R}^{-1}\Big[
	\begin{pmatrix}
	\beta \\
	\beta^*
	\end{pmatrix}
	-
	\bf{S}
	\begin{pmatrix}
	h \\
	h^*
	\end{pmatrix}\Big]\;.
\end{equation}
\noindent Eqs.~(\ref{interN}) and (\ref{Glinear}) set the interaction profile as a function of $\beta$ and $h$. The additional constraint to be taken into account is that the quadratic parameter $\Psi$ should vanish. As shown in Appendix~\ref{Kerrzero}, this amounts to a quadratic equation in $\{h,h^*\}$ which can be solved numerically. 

\section{Gaussian State Reconstruction}\label{sec:Gauss}
%GAUSSIAN STATES AND PARTIAL INFORMATION

A special case of the reconstruction strategy applies when the state $\rho_0$ to be reconstructed is Gaussian. Such states are characterised completely by the first and second order moments of two conjugate quadratures (per mode) and the correlations between them. The collection of these terms is directly accessible to the reconstruction scheme outlined here. In other words, if one knows already that the state is Gaussian, one does not have to reconstruct higher order moments. 

These relevant cases give the opportunity to show explicitly the functioning of the protocol here introduced and to exemplify its requirements. In particular, we will provide examples of the required time-dependent coupling $g(t)$ in units of the mechanical frequency (or smallest eigenfrequency as appropriate), together with estimates of the protocol's performance in terms of fidelity and of the number of measurements necessary to determine the moments associated with each mechanical quadrature. 

\begin{figure}
\includegraphics[width=0.49\columnwidth]{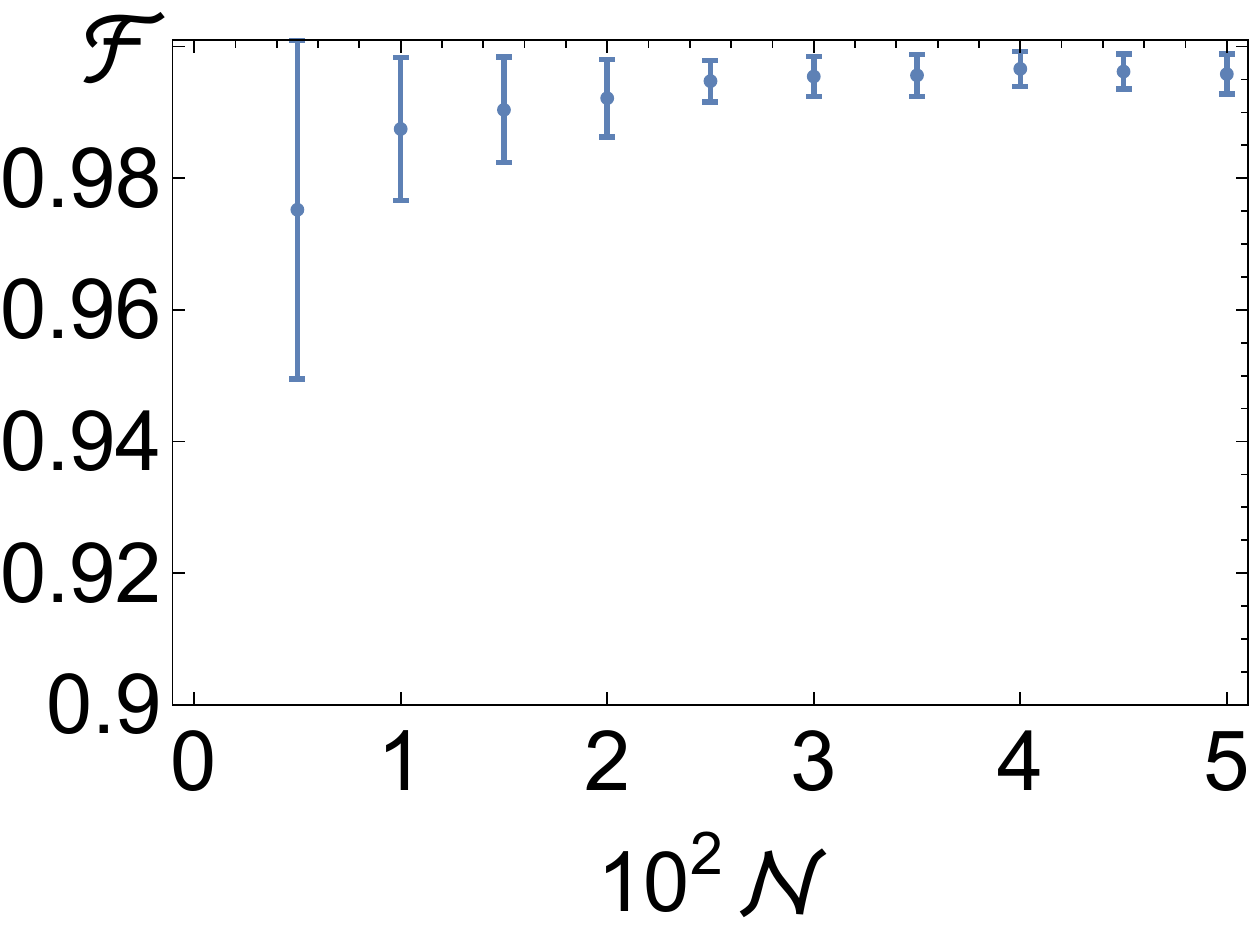}
\includegraphics[width=0.49\columnwidth]{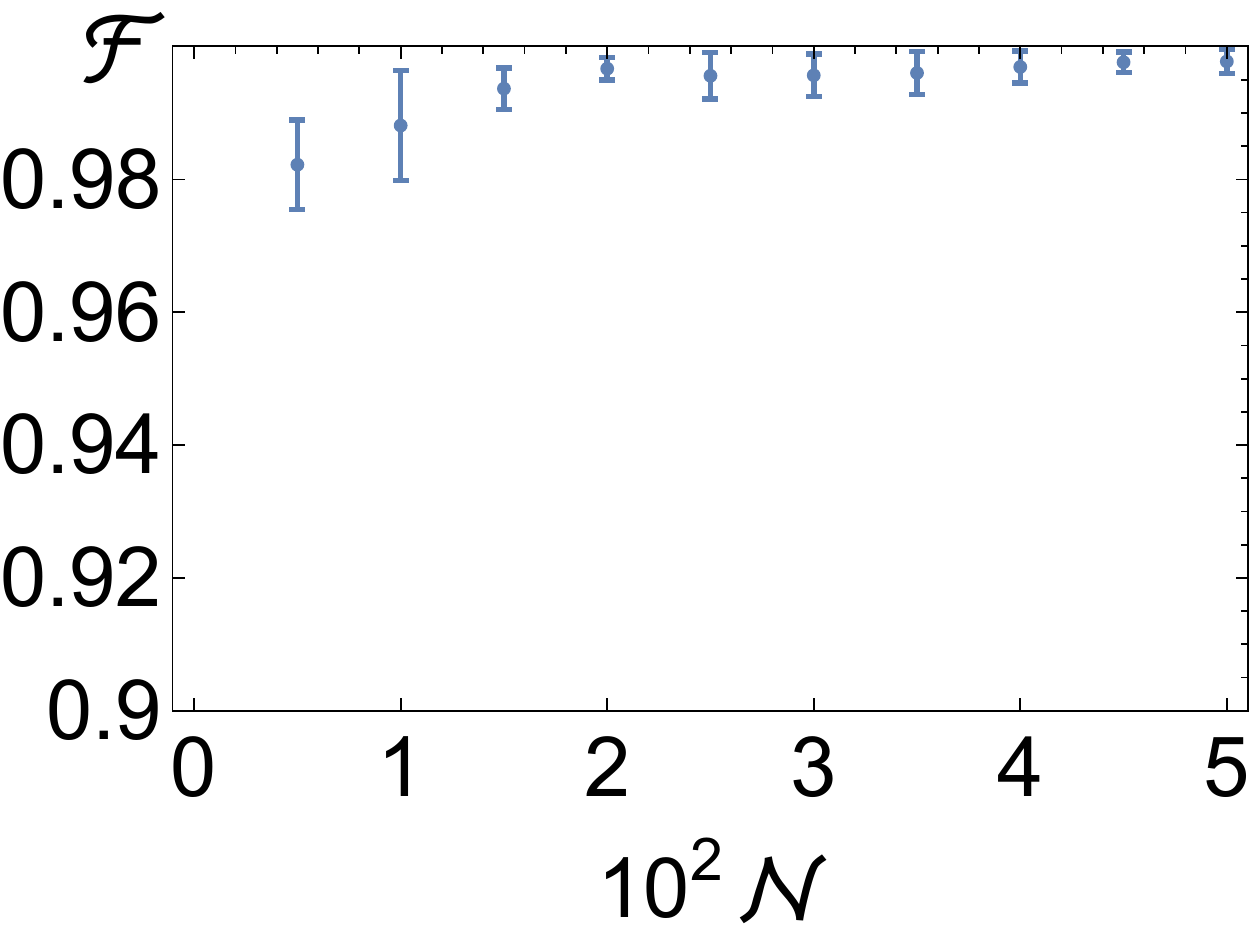}
\caption{The fidelity $\mathcal{F}$ of a single mode thermal state ($T=1$) (left) and a single mode squeezed thermal state ($T=1$, $r=0.2$) (right) against the number of measurements $\mathcal{N}$ in each collection of measurement results determining $\braket{P^n}$. The fidelities are calculated by comparing the reconstructed covariance matrix with the one of $\rho_0$ (see text). In order to exemplify the performance of the reconstruction method, we evaluated the fidelity several times for a fixed $\mathcal{N}$. The points represent the average fidelity thus obtained, whereas the error bars give the respective standard deviation.} \label{Fidelity1}
\end{figure}

\begin{figure}
\includegraphics[width=0.49\columnwidth]{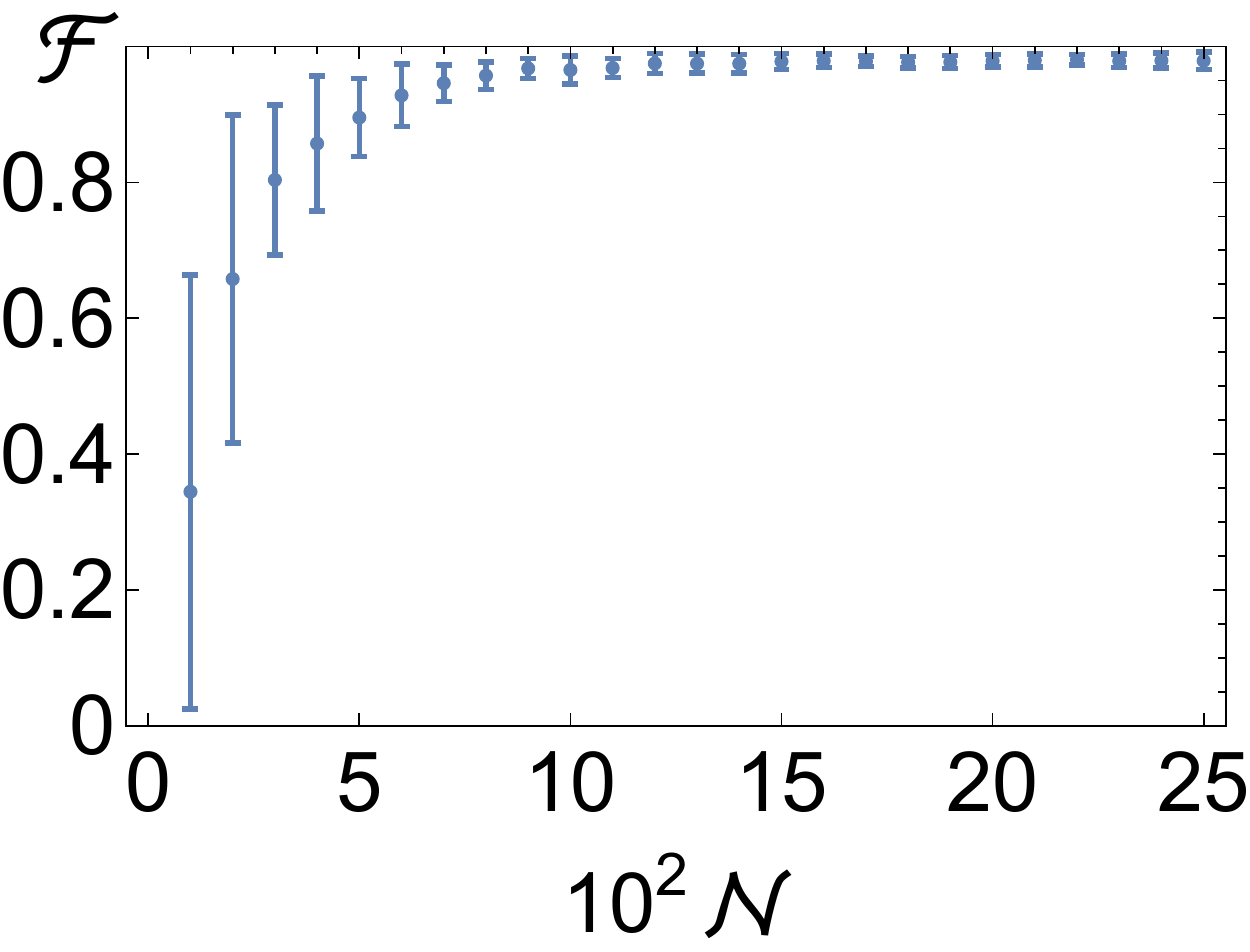}
\includegraphics[width=0.49\columnwidth]{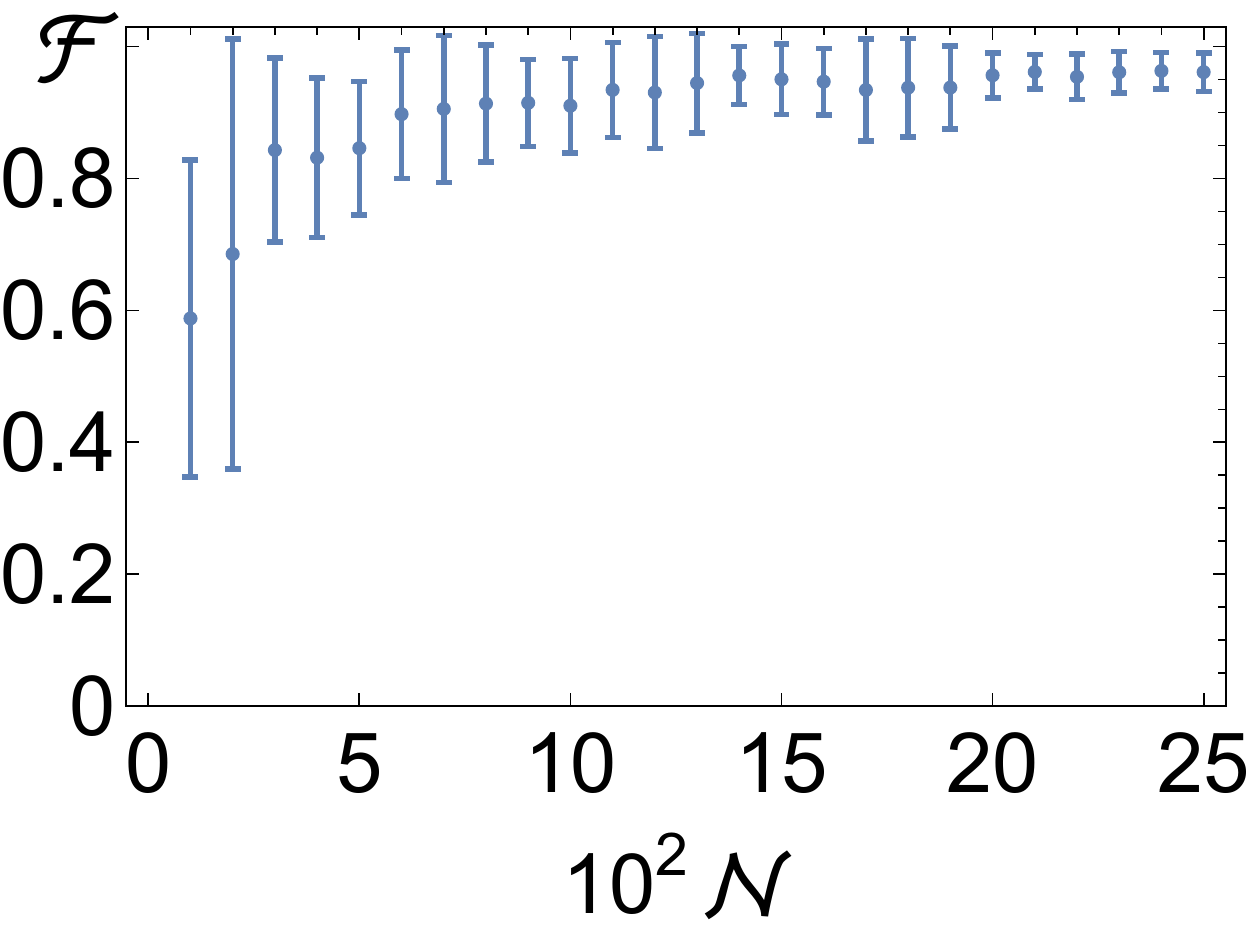}
\caption{The fidelity $\mathcal{F}$ of a two mode thermal state ($T=1.5$) (left) and a two mode squeezed thermal state ($T=1.5$, $r=0.2$) (right) of the normal modes of a network interacting with a spring-like coupling against the number of measurements $\mathcal{N}$ in each collection of measurement results determining $\braket{P^n}$. The fidelities are calculated and illustrated as in Fig.\ref{Fidelity1}} \label{Fidelity2}
\end{figure} 

\begin{figure}
\includegraphics[width=\columnwidth]{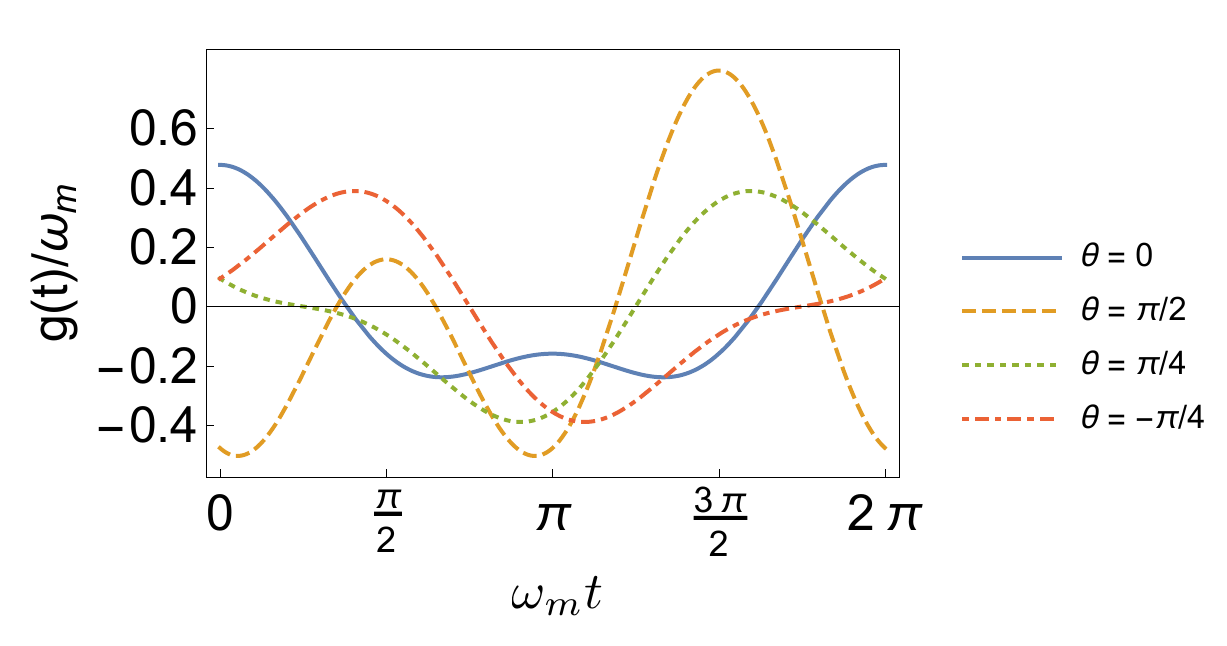}
\caption{Interaction profiles for the case of a single mode reconstruction. The various $\theta$ suffice to reconstruct a single mode Gaussian state.} \label{IntProf}
\end{figure}

\begin{figure}
\includegraphics[width=\columnwidth]{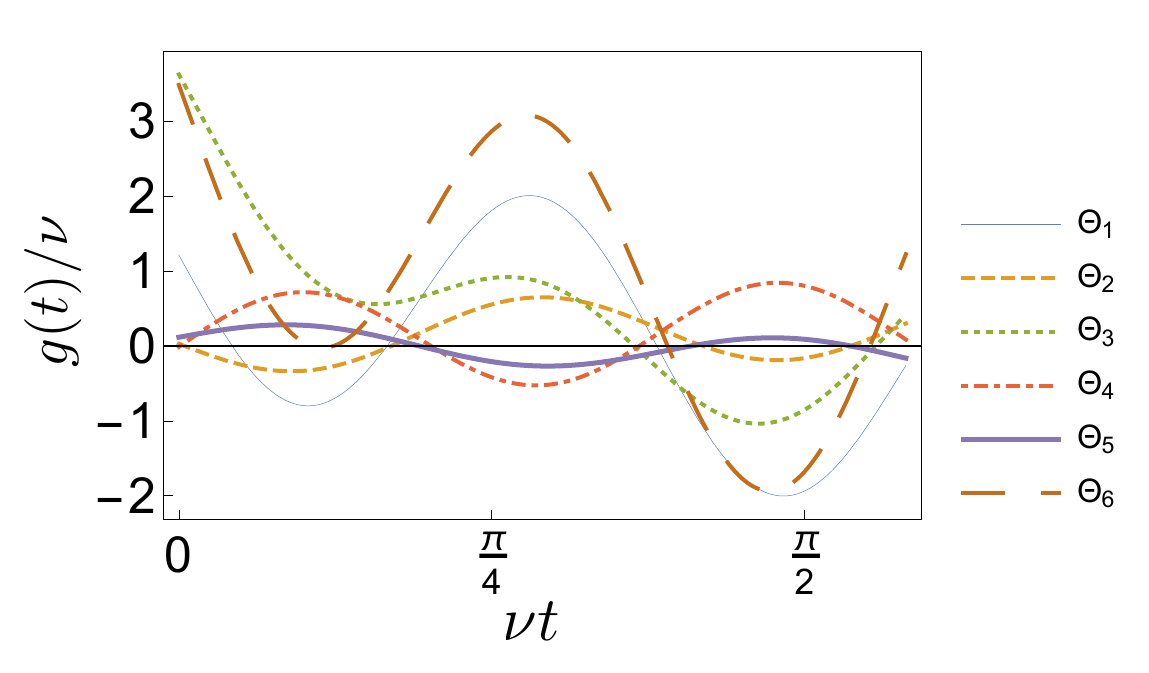}
\caption{Interaction profiles for the case of a two mode reconstruction. The various $\Theta$ suffice to reconstruct a two mode Gaussian state. Each $\Theta$ corresponds to a pair $\{(\theta_1,\theta_2)\}=\{(-\pi/2,-\pi/2)$, $(0,0)$, $(0,-\pi/2)$, $(-\pi/2,0)$, $(-3\pi/4,-3\pi/4)$, $(-\pi/4,-\pi/4)\}$. The parameters of the interaction profiles depend on the choice of interaction times $\tau$. These were chosen independently for each profile $\Theta$ in order to promote their distinguishability. These times were $\tau=5/\nu$, $15/\nu$, $3/\nu$, $25/\nu$, $25/\nu$, $3/\nu$ respectively, where $\nu$ is the smallest eigenfrequency of the two mode network.} \label{IntProf2}
\end{figure}

Figs. \ref{Fidelity1} and \ref{Fidelity2} show the behaviour of the fidelity between the reconstructed state and the original state $\rho_0$. In order to obtain these plots we proceeded as follows. For the single mode case of Fig.~\ref{Fidelity1}, we considered squeezed thermal states with temperature $T$ and squeezing parameter $r$ \cite{Ferraro05}. We set the total interaction time $\tau=2\pi/\omega_m$  and fixed the mechanical quadratures to be reconstructed ($Q_\theta$ with $\theta=0,\pm\pi/4,\pi/2$ suffice in this case). The generic interaction profile is given by \refeq{g_1mode}. The selection of the mechanical quadrature --- together with the additional requirement of deleting the quadratic term $e^{i \psi X^2}$ --- determine as per Eqs.~(\ref{beta},\ref{psi_1}) the specific interaction profile $g(t)$. The latter is reported in Fig.~\ref{IntProf} for the four cases of interest $\theta=0,\pm\pi/4,\pi/2$. Notice that the required tuneability in time is of the order of the mechanical frequency [see also \refeq{g_1mode}] and that the profiles are clearly distinguishable, thus indicating robustness against small perturbations. This range of interaction strengths are typically available experimentally. However, due to experimental limitations, it is also possible that, given a fixed $\beta$, the required magnitude of $g(s)$ is too high. This obstacle can be circumvented by allowing for a longer interaction time, given that the magnitude of $g(s)$ is inversely proportional to it [see Eq.~(\ref{interN})]

For a fixed number of measurements $\cal{N}$, we numerically sampled $\langle Q_{\rm m} \rangle$ and $\langle Q_{\rm m}^2 \rangle$ for the four choices of $Q_\theta$. Then, an inversion of Eq.~(\ref{system1}) allowed for the reconstruction of the first and second moments of $Q_\theta$ from which the covariance matrix of the original state can be reconstructed. We then used the latter and the covariance matrix of $\rho_0$ in order to obtain the fidelity $\cal{F}$ between the reconstructed state and the original one \cite{paraoanu2000fidelity, banchi2015quantum}. Fig.~\ref{Fidelity1} shows that, regardless of the state to be reconstructed, a few hundreds of measurements $\cal{N}$ are sufficient to achieve high fidelity. 

For the two-mode case, we considered two-mode squeezed thermal states. We considered total interaction times $\tau > 1/ \text{min}(\nu_j)$ and the mechanical quadratures given by the set of pairs $\{(\theta_1,\theta_2)\}=\{(-\pi/2,-\pi/2)$,
$(0,0)$, $(0,-\pi/2)$, $(-\pi/2,0)$, $(-3\pi/4,-3\pi/4)$, $(-\pi/4,-\pi/4)\}$. A corresponding set of interaction profiles can be derived using the generic interaction profile of Eq.~(\ref{interN}), and calculating the coefficients following Eq.~(\ref{Glinear}). These profiles have features mirroring those of the single mode case, however the interaction time is not identical for each curve, to improve the distinguishability of each profile (see Fig.~5). The two modes have equal frequencies $\omega=2$ (and therefore distinct eigenfrequencies) and are coupled via a spring-like interaction (Eq.~\ref{linear_net}) with coefficients $J=K=0.7$.

Again, for a fixed number of measurements $\cal{N}$ and for each choice of $\{(\theta_1,\theta_2)\}$, the quadrature $P$ is sampled in order to estimate the second order moments of the mechanical quadratures. In this case, however, extra equations must be generated to make the system in Eq.~(\ref{systemN}) solvable. Each extra equation costs an additional $\cal{N}$ measurements. As can be expected, by increasing the number of oscillators to reconstruct the number of required measurements increases. However, the latter is not significantly affected by the state to be reconstructed.

Clearly, tests of non-Gaussianity are also possible within this scheme. In fact, apart from a full reconstruction of the state as explained in Sections~\ref{sec:oneoscillator} and \ref{sec:network}, one could check the non-Gaussian character of the state $\rho_0$ by reconstructing only few higher-order moments and comparing them with the first and second moments. This is a general feature of this scheme, that it allows direct access to partial information of the state without full tomography.

\section{Single Photon-Phonon Coupling}\label{sec:fottosingle}
%%%%%%%%%%SINGLE PHOTON-PHONON COUPLING

In the previous Sections, we have considered a system in which the radiation-pressure interaction has been treated in the linearized regime [see Eqs.~(\ref{linear_1}) and (\ref{linear_net})]. This is certainly the situation that has been explored most in experiments to date --- both in the opto- and electro-mechanical settings --- giving us a clear motivation to focus on it. However, it is worthwhile to briefly outline how our protocol can be modified to the case in which the radiation-pressure coupling retains its non-linear character. Remarkably, the protocol modifies substantially in this case, and it gives access directly to the characteristic function of the network, as we will now see.

For brevity, we will only mention here the case of one non-linear interacting mechanical resonator $b$. Eq~(\ref{linear_1}) is thus changed into:
\begin{equation}
H=\omega_ca^\dagger a+\omega_mb^\dagger b+g_0a^\dagger a(b+b^\dagger)\;,
\end{equation}
which, in an interaction picture defined by the free terms, becomes:
\begin{equation}
H_I=g_0a^\dagger a(be^{-i\omega_mt}+b^\dagger e^{i\omega_mt})\;.
\end{equation}
The dynamics is solved using the same techniques as above, producing a unitary operator
\begin{equation}
U=e^{i\psi N^2}D(N\beta)\;,
\end{equation}
\noindent where $\psi$ and $\beta$ retain their definitions from before and $N=a^\dagger a$. We can see that now dynamics is described by a number-operator conditioned displacement of the mechanical mode modified by a Kerr-like term on the optical mode. Given a factorised initial state, $\rho=\ket{\alpha}\bra{\alpha}\otimes\rho_0$, where $\ket{\alpha}$ denotes a coherent state, the characteristic function of the mechanical state, $\chi(\beta)=\tr\{D(\beta)\rho_0\}$, may be recovered from the first moments of the cavity position and momentum operators.

Evolving the cavity's position and momentum operators for a time $\tau$ under this displacement operator produces the following relations
\begin{align}
&X(\tau)\equiv D(N\beta)^\dagger XD(N\beta)=X\cosh\hat{\beta}+iP\sinh\hat{\beta}\\
&P(\tau)\equiv D(N\beta)^\dagger PD(N\beta)=P\cosh\hat{\beta}-iX\sinh\hat{\beta}
\end{align}
where $\hat{\beta}=\beta b^\dagger-\beta^*b$ and $X$ and $P$ are defined as before. In this regime of nonlinear coupling we do not have the tuneable coupling required to set $\psi=0$ and therefore the Kerr term cannot be avoided. Including this term, the first moments of $X$ and $P$ are

\begin{multline}
\braket{X}=\bra{\alpha}e^{-i\psi N^2}Xe^{i\psi N^2}\ket{\alpha}\tr(\cosh\hat{\beta})+\\i\bra{\alpha}e^{-i\psi N^2}Pe^{i\psi N^2}\ket{\alpha}\tr(\sinh\hat{\beta})
\end{multline}
\begin{multline}
\braket{P}=\bra{\alpha}e^{-i\psi N^2}Pe^{i\psi N^2}\ket{\alpha}\tr(\cosh\hat{\beta})+\\i\bra{\alpha}e^{-i\psi N^2}Xe^{i\psi N^2}\ket{\alpha}\tr(\sinh\hat{\beta})
\end{multline}

%Given a factorised initial state, $\rho=\ket{\alpha}\bra{\alpha}\otimes\rho_0$, where $\ket{\alpha}$ denotes a coherent state, the characteristic function of the mechanical state, $\chi(\beta)=\tr\{D(\beta)\rho_0\}$, is recovered by direct calculation from
Taking the sum of these produces an expression involving the characteristic function of the mechanical state, $\chi(\beta)=\tr\{D(\beta)\rho_0\}$
\begin{align}
\braket{X}+i\braket{P}&=\bra{\alpha}e^{-i\psi N^2}(X+iP)e^{i\psi N^2}\ket{\alpha}\chi(\beta)\\
&=2\alpha e^{-|\alpha|^2-i\psi}\Big(\sum_n\frac{|\alpha|^{2(n-1)}}{(n-1)!}e^{i2n\psi}\Big)\chi(\beta)
\label{chi_rec}
\end{align}

%Given a factorised initial state, $\rho=\ket{\alpha}\bra{\alpha}\otimes\rho_0$, where $\ket{\alpha}$ denotes a coherent state, the characteristic function of the mechanical state, $\chi(\beta)=\tr\{D(\beta)\rho_0\}$, is recovered by direct calculation from
%\begin{equation}
%\braket{X}+i\braket{P}=2e^{-|\alpha|^2-i\psi}\alpha\Big(\sum_n\frac{|\alpha|^{2(n-1)}}{(n-1)!}e^{i2n\psi}\Big)\chi(\beta)\;,
%\label{chi_rec}
%\end{equation}
%\noindent with $X$ and $P$ defined as before. 

\refeq{chi_rec} above shows a direct link between the expectation values of $X$ and $P$ and the value of the mechanical characteristic function at point $\beta$. By exploring enough points in the phase space it is then in principle possible to reconstruct directly the characteristic function of the mechanical oscillator, which in turn gives full information about its state. This feature is very different with respect to the reconstruction procedure outlined in the previous Sections, and shares a much stricter resemblance with the protocol of \cite{PhysRevA.83.062120,PhysRevA.85.032334}. In general, the direct reconstruction of the characteristic function entails a set of useful features which have been already outlined in the literature. We refer the reader to Refs. \cite{PhysRevA.83.062120,PhysRevA.85.032334} (and references therein) for a detailed account.

We would like to emphasise here that the freedom to explore phase space lies in the definition of $\beta$ [Eq.~(\ref{beta})]. There are two parameters that one could in principle control: the interaction time $\tau$ at which the measurement must be performed and the optomechanical coupling strength $g_0$. A functional dependence on time for $g_0$ would grant the greatest control, but this is difficult to achieve experimentally. Its definition is $g_0=\frac{\omega_c}{L}\sqrt{\frac{\hbar}{2m\omega_m}}$, with $m$ the mass of the oscillator and $L$ the length of the cavity. Since these parameters are usually fixed, the only real freedom is in the interaction time. As is clear from Eq.~(\ref{beta}), changing the interaction time allows exploration of only a ring in phase space, and not the entire space. However, the partial information on the state from this ring may still provide valuable details on the mechanical state.

\section{Conclusions}\label{concussioni}%%%%%%%%%%CONCLUSIONS

We have introduced a method to reconstruct an arbitrary state of a harmonic network of mechanical oscillators. The reconstruction strategy applies to any setting in which a distinguished mechanical oscillator of the network is coupled to a bosonic probe via a linearized interaction. Then, the main feature of our reconstruction protocol is that by measuring a single system (the probe) the state of the entire mechanical network can be recovered. Given that the probe interacts with one mechanical oscillator only, suitable counter-measures can in principle be envisaged in order to screen the rest of the network from sources of noise that are typically unavoidable whenever a system is coupled to a probe. In this sense, our method provides a minimally invasive configuration to monitor a network of oscillators, contrary to a more standard strategy in which each oscillator of the network is individually measured. This is reminiscent of the approach reported in Ref.~\citep{PhysRevA.85.032334}, where a finite dimensional probe was considered. However, in many settings it is more convenient to use an infinite dimensional probe instead. As said, suitable experimental platforms include opto-  and electro-mechanical settings, where linearized coupling at the quantum level has recently been demonstrated between optical or microwave radiation and a single mechanical oscillator \cite{meystre2013short, aspelmeyer2014cavity, rogers2014hybrid}. The main feature that differentiates our setting from the latter is that we consider, rather than only one oscillator, a network of them. However, first implementations of such systems have been reported recently \cite{lin2010coherent, massel2012multimode, shkarin2014optically}, thus providing a promising route towards the realization of small opto- and electro-mechanical networks. In addition, one can show that our protocol can be adapted to configurations in which the mechanical oscillators, rather then being directly coupled, interact only indirectly via a common cavity mode \cite{Bhattacharya2008Optomechanical, Heinrich2011Collective, Chang2011Slowing, Seok2012Optically, Tomadin2012Reservoir, Xuereb2014Reconfigurable}.

In order to assess the performance of our method, we have considered the relevant case in which the state to be reconstructed is Gaussian. In particular, by giving a detailed analysis of one and two-mode cases, we have shown that the quality of the method is oblivious to the details of the reconstructed state. As one could expect, in order for the method to succeed with high fidelity, the number of required measurements increases with the number of modes. We have  explored this feature in some detail by numerically evaluating the fidelity for the realistic case of a finite number of measurements, rather than the limit of an asymptotically large number of measurements as per our analytical results. In addition, we have also shown how the detrimental effect of non-ideal measurements can be taken into account, by considering losses in the coupling between the radiation probe and the mode that is actually measured. Other noise mechanisms could certainly be at work in an actual implementation of our protocol, however they would depend specifically on the platform under consideration and an exhaustive analysis is outside the scope of the present work.

In view of the rapid progress in the development of opto- and electro-mechanical technologies, we believe that the method here introduced could prove useful in assessing the generation of non-classical states of a network of mechanical oscillator, as well as its dynamics. This is of importance in many aspects relevant to the development of quantum technologies, where the quantumness of a system needs to be assessed in detail while minimally compromising its state and its coherent dynamics.

\begin{acknowledgements}
AF and MP acknowledge funding from the John Templeton Foundation (Grant No. 43467). MP thanks the  EU project TherMiQ, the Julian Schwinger Foundation (grant number JSF-14-7-0000), and the UK EPSRC (grants EP/M003019/1). DM acknowledges funding from the EPSRC.
\end{acknowledgements}

\appendix
\section{Effect of genuine optical losses on single oscillator reconstruction} \label{LossesAppendix}
\begin{figure}[H]
\includegraphics[width=\columnwidth]{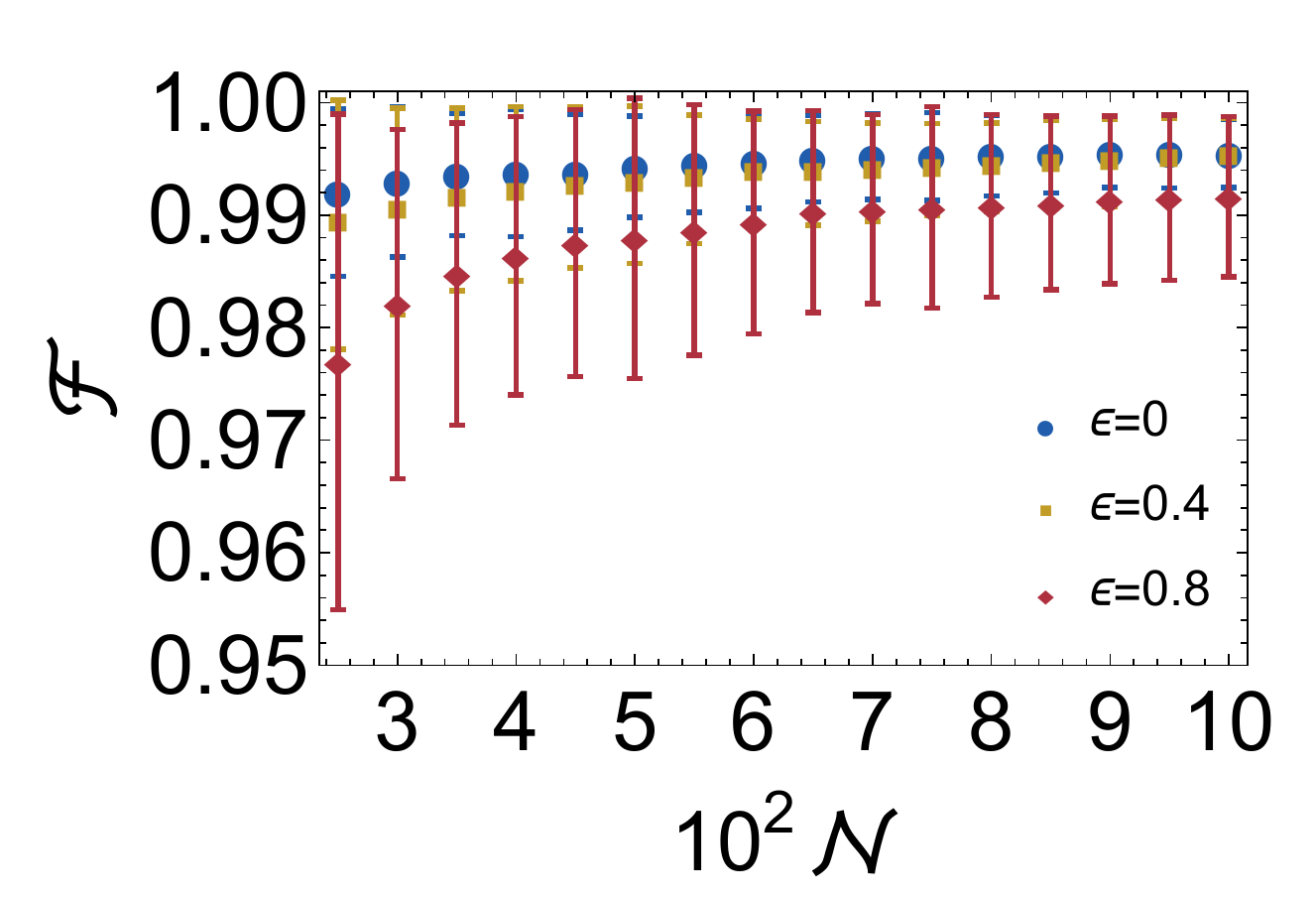}
\includegraphics[width=\columnwidth]{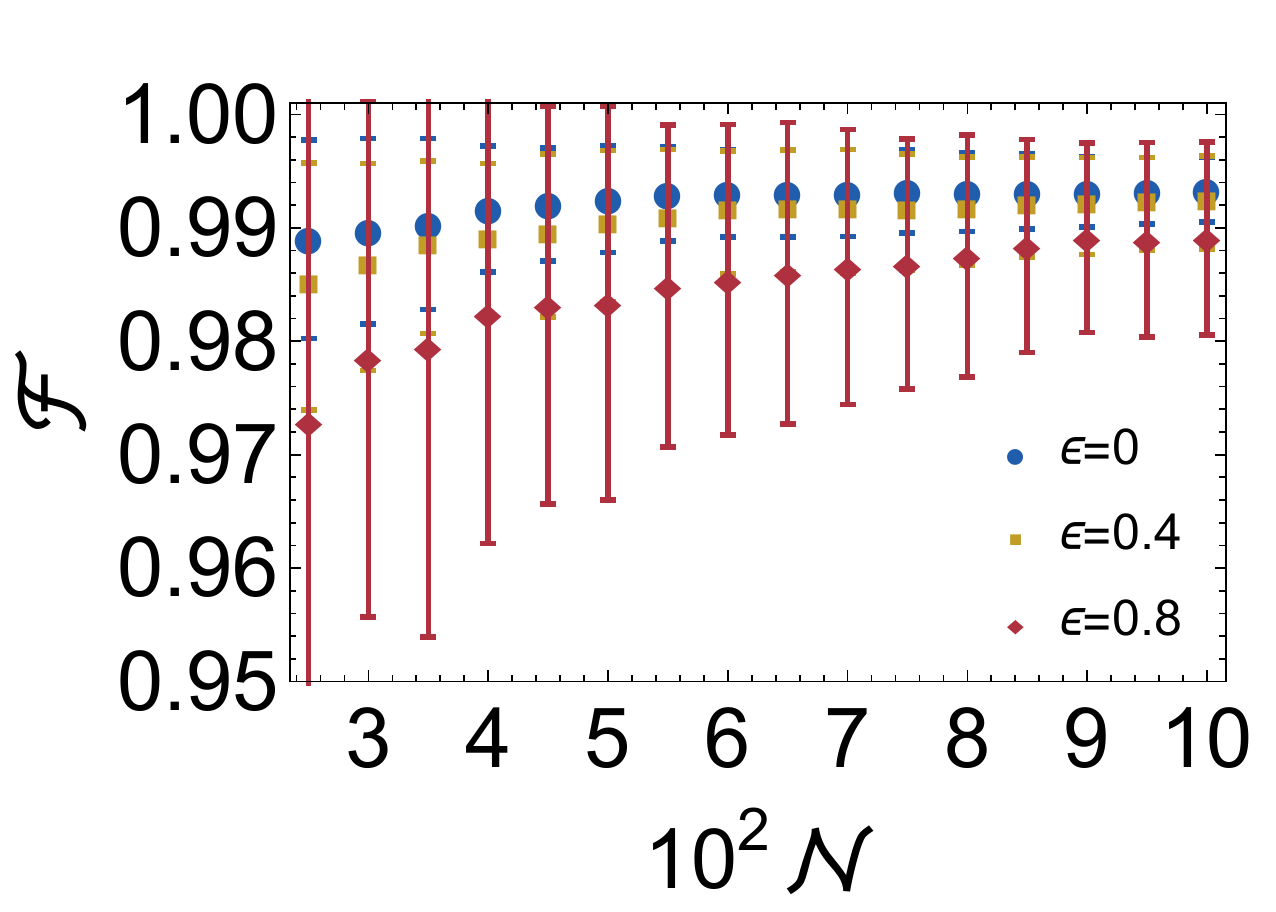}
\caption{The fidelity $\mathcal{F}$ of a single mode thermal state ($T=1$) (left) and a single mode squeezed thermal state ($T=1$, $r=0.2$) (right) against the number of measurement results determining $\braket{P_{\text{out}}^n}$. The fidelities are calculated and illustrated as in Fig.~\ref{Fidelity1}. The calculation is repeated for various values of the genuine optical losses $\epsilon=0,0.4,0.8$, demonstrating that reconstruction is still possible but requires a larger sampling of the observable $P_{\text{out}}$ [see Eq.~(\ref{realistic-readout-moments})].} \label{FidelityLosses}
\end{figure}
In the plots of Fig.~\ref{Fidelity1} it is assumed that the all optical information leaking from the cavity is measurable, or equivalently, that the intracavity field is accessible to direct measurement. In a physical scenario, the losses described in Eq.~(\ref{realistic-readout}) must be taken into account. If the loss coefficient $\epsilon$ is known, then the statistical moments of the intracavity field can be recovered via Eq.~(\ref{realistic-readout-moments}). However, the effect of the losses is to increase the number of measurements (sampling size) required to accurately represent these moments. Fig.~\ref{FidelityLosses} shows the same reconstruction as in Fig.~\ref{Fidelity1} with additional curves demonstrating this effect for various values of $\epsilon$.

\section{Eliminating the quadratic term $e^{i \Psi X^2}$}\label{Kerrzero}
\noindent To write down explicitly the constraint $\Psi=0$, we start by recalling
\begin{equation}
\Psi=\text{Im}\sum_j\int_0^\tau\beta_j\dot{\beta_j}^*dt\;.
\end{equation}
Using the notation of Section~\ref{QSRmulti}, we may perform each integral explicitly
\begin{align*}
\int_0^\tau\beta_j\dot{\beta_j}^*dt=
&\begin{pmatrix}
\mathcal{G} & \mathcal{G}^*
\end{pmatrix}
\begin{pmatrix}
A_j & B_j\\
C_j & D_j
\end{pmatrix}
\begin{pmatrix}
\mathcal{G}\\
\mathcal{G}^*
\end{pmatrix}
\\+
&\begin{pmatrix}
\mathcal{G} & \mathcal{G}^*
\end{pmatrix}
\begin{pmatrix}
E^+_j & E^-_j\\
G^+_j & G^-_j
\end{pmatrix}
\begin{pmatrix}
h\\
h^*
\end{pmatrix}	
\\+
&\begin{pmatrix}
h & h^*
\end{pmatrix}
\begin{pmatrix}
I^+_j & J^-_j\\
J^+_j & I^-_j
\end{pmatrix}
\begin{pmatrix}
\mathcal{G}\\
\mathcal{G}^*
\end{pmatrix}
+
\begin{pmatrix}
h & h^*
\end{pmatrix}
O_j
\begin{pmatrix}
h\\
h^*
\end{pmatrix}
\end{align*}
where
\begin{align}
&A_{nm}=\frac{1}{G_n^*G_m^*}\int_0^\tau e^{-i(\nu_j+\nu_m)t}\int_0^te^{i(\nu_j-\nu_n)s}dsdt\\
&B_{nm}=\frac{1}{G_n^*G_m}\int_0^\tau e^{-i(\nu_j-\nu_m)t}\int_0^te^{i(\nu_j-\nu_n)s}dsdt\\
&C_{nm}=\frac{1}{G_nG_m^*}\int_0^\tau e^{-i(\nu_j+\nu_m)t}\int_0^te^{i(\nu_j+\nu_n)s}dsdt\\
&D_{nm}=\frac{1}{G_nG_m}\int_0^\tau e^{-i(\nu_j-\nu_m)t}\int_0^te^{i(\nu_j+\nu_n)s}dsdt\\
&E^\pm_{n}=\frac{1}{G_n^*}\int_0^\tau e^{\pm i(\nu_j\pm\omega)t}\int_0^te^{i(\nu_j-\nu_n)s}dsdt\\
&G^\pm_{n}=\frac{1}{G_n}\int_0^\tau e^{\pm i(\nu_j\pm\omega)t}\int_0^te^{i(\nu_j+\nu_n)s}dsdt\\
&I^\pm_{n}=\frac{1}{G_n^*}\int_0^\tau e^{-i(\nu_j\pm\nu_n)t}\int_0^te^{i(\nu_j\mp\omega)s}dsdt\\
&J^\pm_{n}=\frac{1}{G_n}\int_0^\tau e^{-i(\nu_j-\nu_n)t}\int_0^te^{i(\nu_j\pm\omega)s}dsdt\;,
\end{align}
and 
\begin{widetext}
	\begin{equation}
	O=
	\begin{pmatrix}
	\int_0^\tau e^{i(\nu_j+\omega)t}\int_0^te^{i(\nu_j-\omega)s}dsdt
	& \int_0^\tau e^{-i(\nu_j-\omega)t}\int_0^te^{i(\nu_j-\omega)s}dsdt\\
	\int_0^\tau e^{-i(\nu_j-\omega)t}\int_0^te^{i(\nu_j+\omega)s}dsdt
	& \int_0^\tau e^{i(\nu_j+\omega)t}\int_0^te^{i(\nu_j+\omega)s}dsdt\\
	\end{pmatrix}
	\end{equation}
\end{widetext}
Finally, we can exploit Eq.~\eqref{Glinear} to substitute the explicit expression for $\mathcal{G}$ as a linear function of $\beta$ and $h$. Thus it is evident that the constraint $\Psi=0$ amounts to a quadratic equation in $\{h,h^*\}$. We have not been able to prove that a solution exists in any instance, however the freedom in the modulus of $\beta_j$ should provide alternatives in case of possible pathological cases. 
%\section{Examples of couplings and phase-space contrast plots} \label{a:details}
%
%This appendix presents some additional details on the reconstruction. First is some further comparison of the effect of number of measurements on the reconstructed Wigner functions. This can be seen by taking covariance matrices corresponding to points from plots in Fig. \ref{Fidelity} and looking at the pointwise absolute difference between their Wigner functions and that of the reference state. Going left to right in Fig. \ref{Wig} indicates that an decreased number of measurements increases the difference between these states.
%
%\begin{figure}[h]
%\includegraphics[width=0.4\columnwidth]{WigBad.pdf}
%\includegraphics[width=0.5\columnwidth]{WigUgly.pdf}
%\caption{Absolute values of pointwise differences between reconstructed Wigner functions with different fidelities (left to right: $\mathcal{F}=0.99$, $\mathcal{F}=0.82$; number of measurements per run $2000$ and $10$ respectively) and the ideal Wigner function of a squeezed thermal state of Fig. \ref{Fidelity}b. } \label{Wig}
%\end{figure}
\bibliography{Biblio.bib}
\end{document}